\documentclass[useAMS,aastex]{emulateapj}
\def\approxgt{\ifmmode \rlap{$>$}{}_{{}_{{}_{\textstyle\sim}}} \else%
$\rlap{$>$}{}_{{}_{{}_{\textstyle\sim}}}$\fi} 
\def\approxlt{\ifmmode \rlap{$<$}{}_{{}_{{}_{\textstyle\sim}}} \else%
$\rlap{$<$}{}_{{}_{{}_{\textstyle\sim}}}$\fi}

\def\arcsec{\hbox{$^{\prime\prime}$}}

\def\flx{erg cm$^{-2}$ s$^{-1}$}
\def\lum{erg s$^{-1}$}

\def\chan{{\it Chandra}}
\def\src{MAXI~J1659--152}
\def\swift{{\it Swift}}
\newcommand{\ledd}{$L_{Edd}$}

\shorttitle{The outburst decay of \src}
\shortauthors{Jonker et al.}

\begin{document}

\title{The black hole
  candidate \src\, in and towards quiescence in X--ray and radio }

\author{P.G.~Jonker\altaffilmark{1,2,3}}
\email{p.jonker@sron.nl}

\author{J.C.A.~Miller--Jones\altaffilmark{4}} 
\author{J.~Homan\altaffilmark{5}} 
\author{J.~Tomsick\altaffilmark{6}}
\author{R.P.~Fender\altaffilmark{7}} 
\author{P.~Kaaret\altaffilmark{8}}
\author{S.~Markoff\altaffilmark{9}} 
\author{E.~Gallo\altaffilmark{10}}
\altaffiltext{1}{SRON, Netherlands Institute for Space Research,
Sorbonnelaan 2, 3584~CA, Utrecht, The Netherlands}
\altaffiltext{2}{Harvard--Smithsonian Center for Astrophysics, 60 Garden Street, Cambridge, MA~02138, U.S.A.}
\altaffiltext{3}{Department of Astrophysics/IMAPP, Radboud University Nijmegen,
P.O. Box 9010, 6500 GL Nijmegen, The Netherlands} 
\altaffiltext{4}{International Centre for Radio Astronomy Research, Curtin University, GPO Box U1987, Perth, WA 6845, Australia}
\altaffiltext{5}{MIT, Kavli Institute for Astrophysics and Space Research, 70 Vassar Street, Cambridge, MA 02139, USA}
\altaffiltext{6}{Space Sciences Laboratory, University of California, Berkeley, USA}
\altaffiltext{7}{School of Physics and Astronomy, University of Southampton, Southampton SO17 1BJ}
\altaffiltext{8}{Department of Physics and Astronomy, University of Iowa, Van Allen Hall, Iowa City, IA 52242, USA}
\altaffiltext{9}{Astronomical Institute `Anton Pannekoek', University of Amsterdam, Postbus 94249, 1090 GE Amsterdam, the Netherlands}
\altaffiltext{10}{Department of Astronomy, University of Michigan, 500 Church Street, Ann Arbor, MI 48109, USA}


\begin{abstract} \noindent In this paper we report on Expanded Very
  Large Array radio and \chan~and \swift~X--ray observations of the
  outburst decay of the transient black hole candidate \src\, in 2011.
  We discuss the distance to the source taking the high inclination
  into account and we conclude that the source distance is probably
  6$\pm2$ kpc. The lowest observed flux corresponds to a luminosity of
  $2\times 10^{31}(\frac{{\rm d}}{6 {\rm kpc}})^2$\lum. This, together
  with the orbital period of 2.4~hr reported in the literature,
  suggests that the quiescent X--ray luminosity is higher than
  predicted on the basis of the orbital period -- quiescent X--ray
  luminosity relationship. It is more in line with that expected for a
  neutron star, although the outburst spectral and timing properties
  reported in the literature strongly suggest that \src\, harbors a
  black hole. This conclusion is subject to confirmation of the
    lowest observed flux as the quiescent flux. The relation between
  the accretion and ejection mechanisms can be studied using the
  observed correlation between the radio and X--ray luminosities as
  these evolve over an outburst.  We determine the behaviour of \src\,
  in the radio -- X--ray diagram at low X--ray luminosities using the
  observations reported in this paper and at high X--ray luminosities
  using values reported in the literature. At high X-ray luminosities
  the source lies closer to the sources that follow a correlation
  index steeper than 0.6--0.7.  However, when compared to other
  sources that follow a steeper correlation index, the X--ray
  luminosity in \src\, is also lower.  The latter can potentially be
  explained by the high inclination of \src\, if the X--ray emission
  comes from close to the source and the radio emission is originating
  in a more extended region. However, it is probable that the
    source was not in the canonical low--hard state during these radio
    observations and this may affect the behaviour of the source as
    well. At intermediate X--ray luminosities the source makes the
  transition from the radio underluminous sources in the direction of
  the relation traced by the 'standard' correlation similar to what
  has been reported for H~1743--322 in the literature.  However,
  \src\, remains underluminous with respect to this 'standard'
  correlation.
\end{abstract}

\keywords{stars: individual (\src) --- accretion: accretion discs --- stars: binaries 
--- X-rays: binaries}

\section{Introduction} 

Nearly all black hole X--ray binaries (BHXBs) are found in transient
sources that show a dramatic increase in their X--ray luminosity (see
\citealt{2006csxs.book..157M} for a review). Such transient BHXBs
spend long periods at very low X--ray luminosities, referred to as
`quiescence'. The luminosity increases during occasional outbursts by
as much as 7 to 8 orders of magnitude, typically reaching values of
tens of per cent of the Eddington luminosity (\ledd). The X--ray
spectral and timing properties of BHXBs evolve in a characteristic way
as the outburst progresses
(e.g.~\citealt{2010AIPC.1248..107B}). Furthermore, the X--ray behaviour
is a good predictor of the radio behaviour suggesting an intimate
connection between the two (e.g.~\citealt{2009MNRAS.396.1370F}).

\citet{1998A&A...337..460H} and later \citet{2003A&A...400.1007C} and
\citet{2003MNRAS.344...60G} found that the X--ray and radio luminosity
from hard state black holes are correlated.  The correlation takes the
form $L_{\rm radio} \propto L_{\rm X}^{0.6-0.7}$.  However, since then
more data has become available for other sources
(e.g.~\citealt{2007A&A...466.1053X}, \citealt{2007ApJ...659..549C},
\citealt{2007ApJ...655L..97R}, \citealt{2007ApJ...655..434S},
\citealt{2008MNRAS.389.1697C}, \citealt{2010MNRAS.401.1255J},
\citealt{2011MNRAS.414..677C}, Ratti et al.~2012 accepted) that
revealed that several sources follow a steeper correlation between the
X--ray and radio luminosity. Recently,
\citet{2011MNRAS.414..677C} found that the BHXB H~1743--322 followed
the sources on this 'outlier' branch at X--ray luminosities above
$2\times 10^{36}$ \lum, but when the source luminosity decreased it
showed a transition at an approximately constant radio luminosity of a
few $\times 10^{19}$ \lum${\rm Hz^{-1}}$ to the 'standard' power--law
correlation with index 0.6. The source XTE~J1752--223 shows behaviour
consistent with the same trend (Ratti et al.~submitted).

\citet{2005ApJ...624..295H} found a similar correlation between the
X--ray and near--infrared flux in the BHXB GX~339--4.
\citet{2006MNRAS.371.1334R} used a much larger sample of sources and
found a similar correlation between the X--ray and near--infrared flux
in all the BHXBs in the low--hard state that they investigated:
$L_{IR} \propto L_X^{\,\sim0.6}$. It is at present not clear to what
extend the near--infrared -- X--ray correlation is caused by the jet
outflow. Nevertheless, besides the potential connection between the
radio and X--ray luminosity, the X--ray and near--infrared luminosity
in the BHXBs in the hard state are also connected. So far, it has been
possible to test the radio/X--ray relation in quiscence only with
quasi--simultaneous observations for A~0620--00;
  (\citealt{2006MNRAS.370.1351G}) and for V404~Cyg
  (\citealt{2009MNRAS.399.2239H}). It was found that the radio and
X--ray flux lie on the extension of the $L_{\rm radio} \propto L_{\rm
  X}^{0.7}$ 'standard' correlation, suggesting that it holds all the
way down to quiescence. This seems to rule out models that predict
that the relation should significantly steepen around
$10^{-5}$\,\ledd\ (\citealt{2005ApJ...629..408Y}).

Here, we present a study of \src, it was discovered by the Monitor of
All-sky X--ray Image instrument onboard the International Space
Station (MAXI; \citealt{2009PASJ...61..999M}). Shortly thereafter it
was detected by the \swift\, Burst Alert Telescope (BAT;
\citealt{2010GCN..11296...1M}; note that the BAT detection was
reported earlier than the MAXI detection). Initially it was classified
as a gamma--ray burst (GRB~100925A), but it was quickly realized that
the source is a strong black hole candidate (BHC;
\citealt{2010ATel.2881....1K}).

Dips in the X--ray light curve of \src\,reveal the orbital period. At
a period of 2.4 hrs (\citealt{2010ATel.2912....1K};
\citealt{2011ApJ...736...22K}) this is currently the shortest orbital period
known for a BHXB. The X--ray timing and broad--band spectral
properties evolve over the outburst in a canonical way
(\citealt{2011ApJ...736...22K}; \citealt{2011ApJ...731L...2K};
\citealt{2011MNRAS.415..292M}) strengthening the identification of the
source as a black hole candidate(BHC). The faintness of the optical
counterpart in quiescence (\citealt{2010ATel.2976....1K}) and the
short orbital period will make it difficult to obtain an accurate mass
measurement through time--resolved optical spectroscopy of the companion
star while the system is in quiescence. \citet{2011ApJ...736...22K}
constrain the distance to the source to be more than 6.1 kpc.  An
accurate source position was obtained using radio very long baseline
interferometry (VLBI) observations using the European VLBI Network
(EVN; \citealt{2010ATel.2906....1P}).

Here, we report on contemporary \chan, \swift\, X--ray and Expanded
Very Large Array (EVLA) radio observations of the black hole candidate
\src\, aimed at following the X--ray and radio light curves and
establishing the X--ray -- radio correlation during the last part of
the decay to quiescence as well as the X--ray luminosity in
quiescence.
 
\section{Observations, analysis and results} 

\subsection{\chan\, X--ray data}
We observed \src\, with the \chan\, satellite using the
back--illuminated S3 CCD--chip of the Advanced CCD Imaging
Spectrometer (ACIS) detector (\citealt{1997AAS...190.3404G}) on six
occasions during the decay to quiescence (see Table~\ref{chanlog} for
a journal of the \chan\, observations).  We windowed the ACIS--S CCD
in all observations, providing a frame time of 0.4104~s. We have
reprocessed and analysed the data using the {\it CIAO 4.3} software
developed by the Chandra X--ray Center employing the calibration files
from the Calibration Database version 4.4.6. The last observation (ID
12443) has been observed with the datamode set to {\sc VFAINT}. This
means that pulse height information in a 5x5 pixel region around the
event is telemetered down, allowing for a more rigorous cleaning of
background events caused by for instance cosmic rays. In our analysis
we have selected events only if their energy falls in the 0.3--7 keV
range.  All data have been used, as background flaring is very weak or
absent in all observations. For the last two observations (ID 12442
and 12443) we produced custom--made bad pixel maps taking into account
the change in the \chan\, aimpoint on the ACIS CCD.

\begin{table*}
\caption{A journal of the \chan\ observations. The MJD and the observing date refer to the start time of the observation.}
\label{chanlog}
\begin{center}
\begin{tabular}{cccccc}
\hline
Obs ID & Observing  & MJD & Time on & Net count rate & {\sc wavdetect} detected \# \\
& date & (days; UTC) & source (ks) & 0.3-7 keV (cnt s$^{-1}$) & source counts \\
\hline\hline
12438$^a$ & 2011 Apr.~14 & 55665.96202 & 6.36 & $(3.9\pm0.3)\times10^{-2}$ & 235 \\
12439$^a$ & 2011 Apr.~23 & 55674.74944 & 9.08 & $(9.9\pm1.1)\times10^{-3}$ & 80 \\
12440 & 2011 May~03 & 55684.29844 & 13.6 & $(4.8\pm3.5)\times10^{-4}$ & 8 \\
12441$^a$ & 2011 May~12 & 55693.21054 & 18.1 & $(6.87\pm0.06)\times10^{-1}$ & 11761 \\
12442$^a$ & 2011 Aug.~15 & 55788.83283 & 30.8 & $(5.5\pm2.3)\times10^{-4}$ & 10 \\
12443 & 2011 Oct.~12 & 55846.53179 & 90.7 & $(4.5\pm1.3)\times10^{-4}$ & 39 \\

\end{tabular}		      
\end{center}
{\footnotesize $^a$ This observation is used in Figure~\ref{xrcorr}.\\}

\end{table*}

\begin{figure*} 
\includegraphics[angle=0,height=6cm,clip]{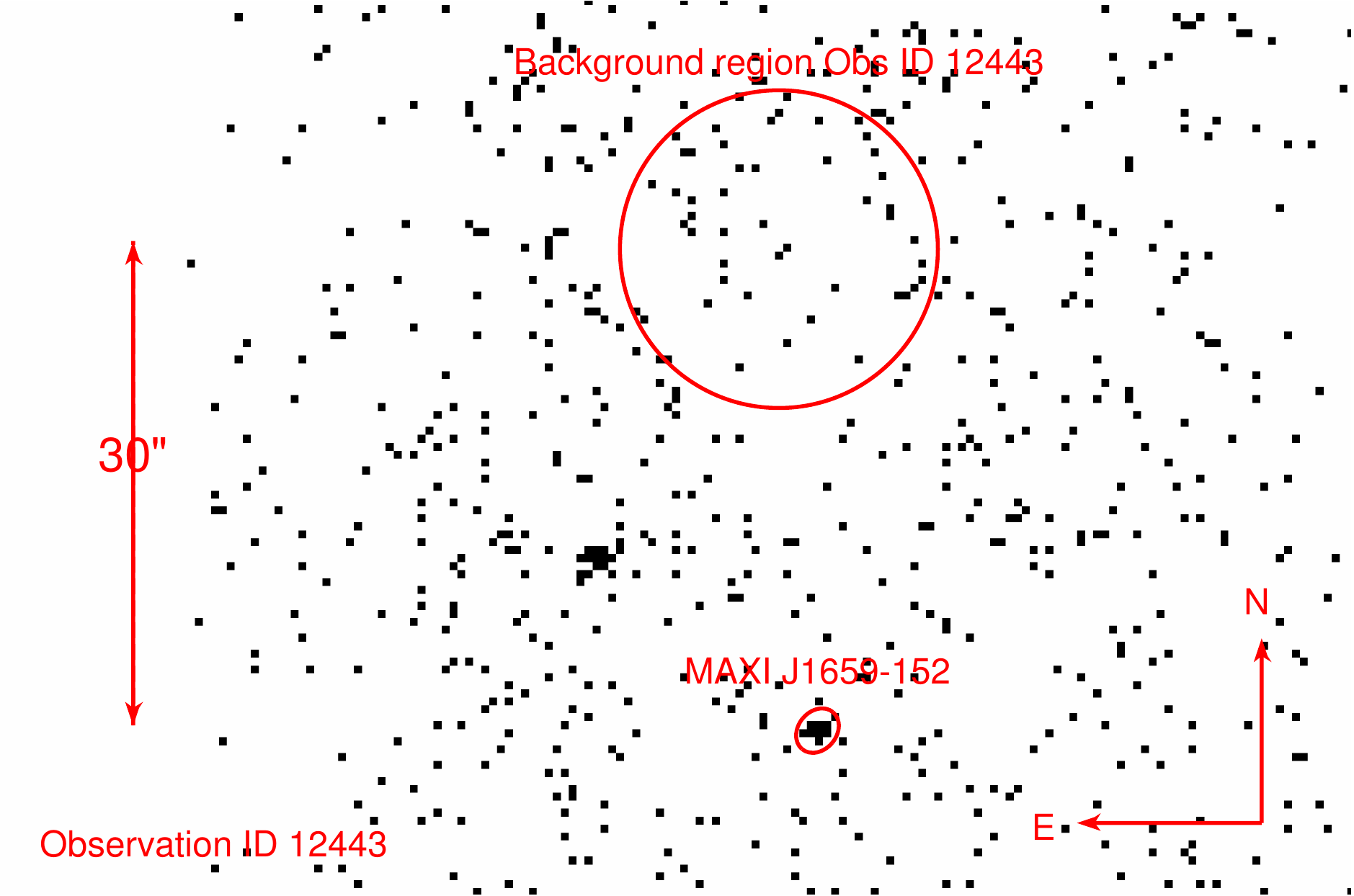}
\includegraphics[angle=0,width=8cm,clip]{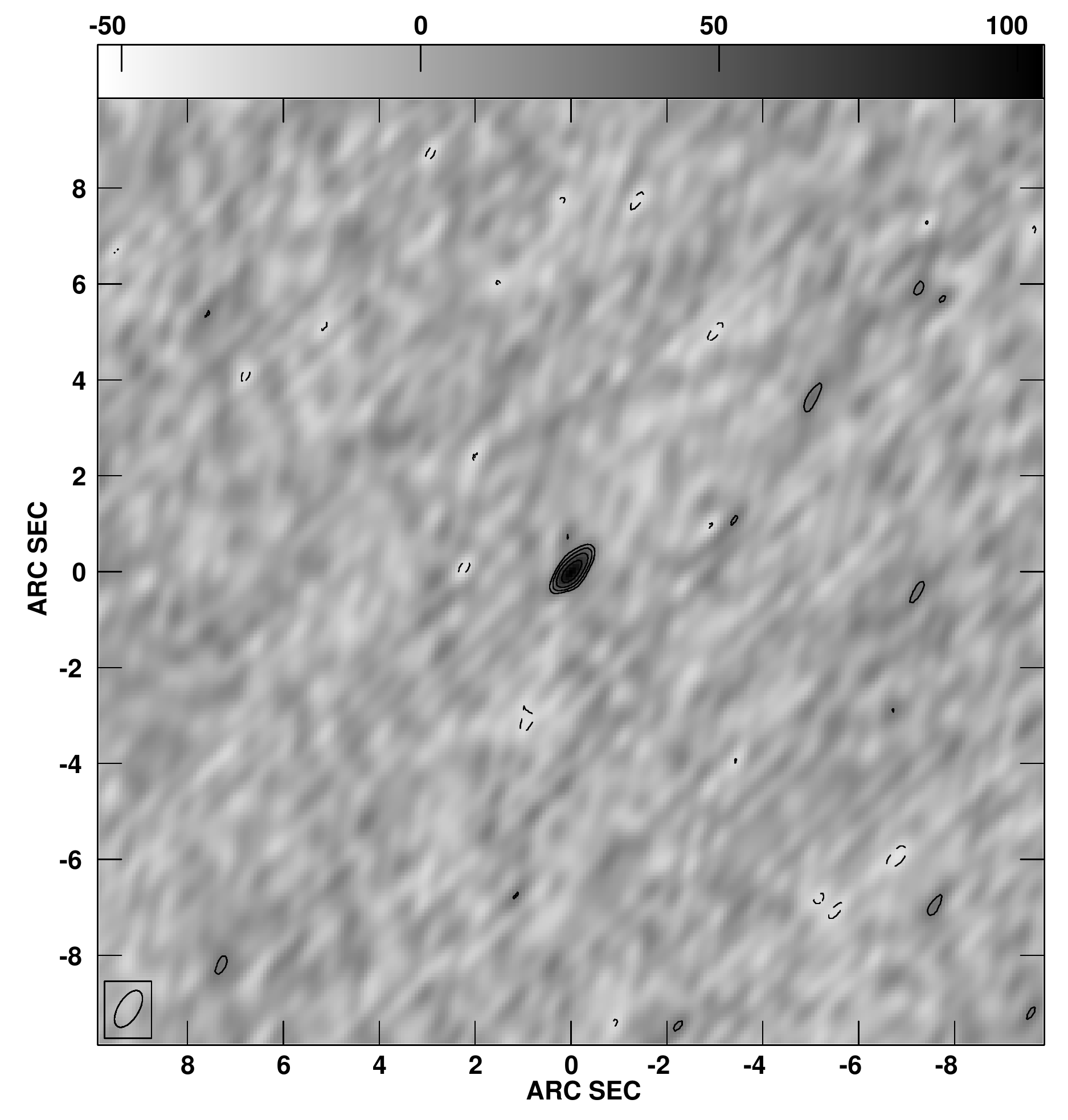}
\caption{{\it Left panel:} The X--ray 0.3--7 keV {\it Chandra} image
  of Obs ID 12443 indicating \src\, and the region used for estimating
  the background. {\it Right panel:} The EVLA radio image at a
  frequency of 4.96 GHz on 2011 June 29 of dimension 10\arcsec\, on a
  side, with \src\, at the centre.  The colour bar is in microJy/beam.
  The contour levels are at levels of +/-sqrt(2)$^n$ times the rms
  noise of 10 microJy/beam, where n=3,4,5,..}
\label{lc} \end{figure*}

Using {\sl wavdetect} we detect \src\ in each of the observations. We have
selected a circular region of 10\arcsec\,radius centered on the accurately known
source position (\citealt{2010ATel.2906....1P}) to extract the source counts. 
Similarly, for the first five \chan\, observations, we have used a circular
region with a radius of 20\arcsec\, on a source--free region of the CCD to
extract background counts. For the sixth and last observation with observation
ID 12443 we used a circular region with a radius of 10\arcsec\, as the long
integration revealed some weak point sources in the region covered by the
20\arcsec\,--radius background circle. Nevertheless, as the exposure is more
than 90 ksec.~long, there are sufficient counts in this circle in order to
reliably estimate the background event rate.

The redistribution response file is the same for the source and
background region but we have made auxilliary response matrices for
the source region of each of the observations separately. The net,
background subtracted, source count rate for each observation is given
in Table~\ref{chanlog}.

Using {\sl xspec} version 12.6.0q (\citealt{ar1996}) we have fitted
the spectra of \src\ using Cash statistics
(\citealt{1979ApJ...228..939C}) modified to account for the
subtraction of background counts, the so called
W--statistics\footnote{see
  http://heasarc.gsfc.nasa.gov/docs/xanadu/xspec/manual/} for all six
observations. We have used an absorbed power--law model ({\sl
  pegpwrlw} in {\sl xspec}) to describe the data.
  
\citet{2011ApJ...736...22K} found evidence for variations in ${\rm
  N_H}$ as the outburst progressed. Initially they found ${\rm
  N_H}=$2.3$\pm$0.3$\times 10^{21}$ cm$^{-2}$ whereas later during the
outburst an average value of 5$\times 10^{21}$ cm$^{-2}$ was found (no
error bar was given for the latter value). Given that in most
observations we only detected a low number of counts we decided to fix
the ${\rm N_H}$ to 2.3$\times 10^{21}$ cm$^{-2}$ in our spectral
fits. If a significant amount of this extra (local) absorbing material
was still present during the time of our observations we will slightly
under estimate the value of the power--law index in our fits.  Owing
to the high count rate during observation ID 12441, we included the
pile--up model (\citealt{2001ApJ...562..575D}) in the fit. For
reference we also performed the fit without the pile--up model. The
difference between the two fits is small. Finally, we fitted the
spectrum of observation ID 12441 leaving the ${\rm N_H}$ parameter
free as well (including or excluding the pile--up model has no
significant effect on the best--fit value for ${\rm N_H}$). We found a
best fit value ${\rm N_H}=(2.7\pm0.1)\times 10^{21}$ cm$^{-2}$.  We
list the results of our spectral analysis in Table~\ref{spec}. See
Table~\ref{val-xrcorr} for the 1--10 keV flux values that we derived.

\begin{table*}
\caption{Best fit parameters of the spectra of \src. PL refers to
  power law.  All quoted errors are at the 68 per cent confidence
  level. }
\label{spec}
\begin{center}
\begin{tabular}{ccccc}
\hline
Obs  & PL  & Unabs.~0.5--10 keV & Abs.~0.5--10 keV  & Goodness \\ 
 ID &   index &  erg$^{-1}$~cm$^{-2}$~s$^{-1}$ &  erg$^{-1}$~cm$^{-2}$~s$^{-1}$ & per cent \\
\hline
\hline
12438 &  2.1$\pm$0.1 & $(4.0\pm0.3)\times 10^{-13}$  & $(3.0\pm0.4)\times 10^{-13}$  & 60 \\
12439 &  2.5$\pm$0.3 & $(7\pm1)\times 10^{-14}$  & $(5\pm1)\times 10^{-14}$  & 99 \\
12440 &  2$^a$ & $(9\pm3)\times 10^{-15}$  & $(7\pm2)\times 10^{-15}$  & 26 \\
12441 &  1.46$\pm$0.02 & $(9.4\pm0.1)\times 10^{-12}$  & $(8.2\pm0.1)\times 10^{-12}$  & 89 \\
12441$^b$ &  1.48$\pm$0.03 & $(1.9\pm0.2)\times 10^{-11}$  & $1.7\times 10^{-11}$  & 89 \\
12442 &  2$^a$ & $(6\pm2)\times 10^{-15}$  & $(4.6\pm1.0)\times 10^{-15}$  & 67 \\
12443 &  3.2$\pm$0.6 & $(4\pm1)\times 10^{-15}$  & $(2.3\pm0.7)\times 10^{-15}$  & 75 \\
\hline
\end{tabular}
\end{center}
{\footnotesize $^a$ Parameter fixed at this value. \\}
{\footnotesize $^b$ Fit--function includes the pile-up model, $\alpha=0.03^{+0.09}_{-0.03}$ (\citealt{2001ApJ...562..575D}). \\}
\end{table*}

\subsection{Selected \swift~XRT observations}

We analyse selected \swift\, X--ray telescope (XRT) observations that
were obtained close in time to our EVLA observations in order to track
the radio -- X--ray correlation. The \swift\, data were reduced using
dedicated tools within {\sc heasoft } version 6.11. We applied
filtering based on the standard events grades (0--12 in photon
counting mode). The calibration database version 4.4 was used. We
employed the tools {\sc xrtpipeline} and {\sc xrtproducts} to obtain
the XRT spectra. We fitted the spectra using {\sc xspec} following the
same procedure as for the \chan\, spectra.  We plot the unabsorbed
0.5--10 keV X--ray flux obtained from the \swift\, observations with
IDs 00031843014--00031843025 in Figure~\ref{lc}. We excluded the
observations with IDs 00031843016 and 00031843017 from our analysis as
there are no data products suitable for our purpose. For the radio --
X--ray correlation (see below) we calculated 1--10 keV unabsorbed
fluxes and we used the \swift\, observation on MJD~55721.22 (Obs.~ID
00031843018) and MJD 55741.02 (Obs.~ID 00031843023) as they are close
in time to our EVLA observations on 2011 Jun.~03 (MJD~55715.21) and
2011 Jun.~29 (MJD~55741.14), respectively. See Table~\ref{val-xrcorr}
for the \swift\, 1--10 keV flux values that we derived.

In the {\it top panel } of Figure~\ref{lc} we show the \swift\, and
\chan\, light curve of the last phase of the 2010--2011 outburst of
\src. After an initial decay to (close to) quiescence, a reflare
occurred after MJD~55684 (May 03, 2011), where the source brightened
by a factor of 180 in less than 3 days. Next, the source remained
brighter than 1$\times 10^{-12}$\flx\, for a period of about 2 months
before it decayed to the quiescent flux level of 5$\times
10^{-15}$\flx\, in less than 40 days.

\begin{figure} 
\includegraphics[angle=0,width=8cm,clip]{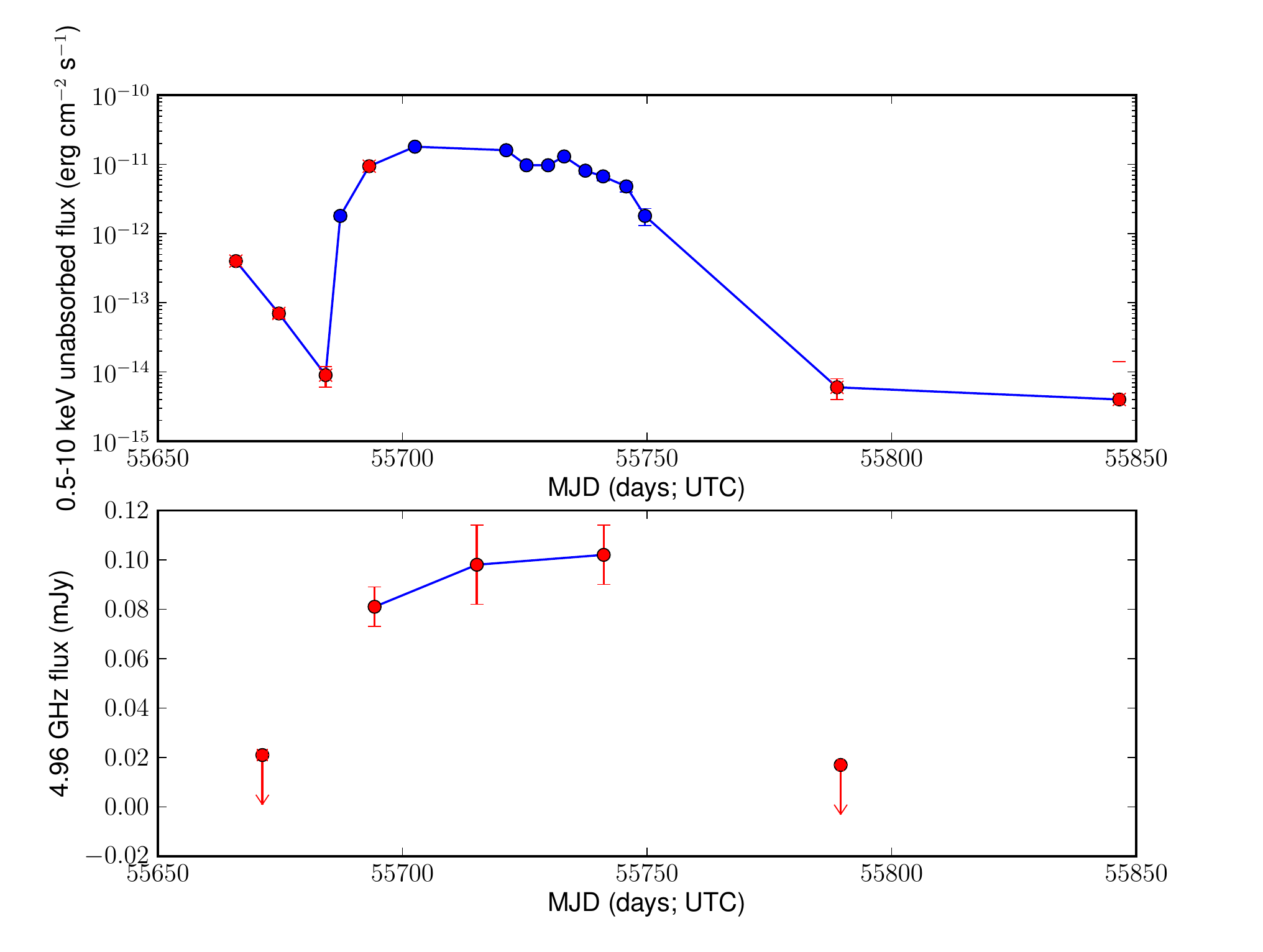}
\caption{{\it Top panel:} The X--ray 0.5--10 keV light curve derived
  from our {\it Chandra} (red symbols) and public \swift\, (blue
  symbols) monitoring during the decay towards quiescence after the
  2011 outburst of \src. After MJD 55684 the source exhibited a
  reflare before decaying towards quiescence.  {\it Bottom panel:} The
  quasi--simultaneous EVLA radio monitoring of the source at a
  frequency of 4.96 GHz. }
\label{lc} \end{figure}

\subsection{EVLA radio observations}

\label{sec:evla}

We observed MAXI J1659--152 with the Expanded Very Large Array
\citep[EVLA;][]{2011ApJ...739L...1P} over 7 epochs from 2011 April 19 through 
2011 August 17,
under program code SC0346.  The array was initially in the relatively extended
`B' configuration, had moved to the hybrid `BnA' configuration prior to our
observations on 2011 May 13, and by 2011 June 29 was in its most extended `A'
configuration.  We observed at a central frequency of 4.96\,GHz, using 256\,MHz
of contiguous bandwidth split into two 128-MHz sub-bands, each of which
comprised sixty-four 2-MHz channels.

The data were reduced using the Common Astronomy Software Application
\citep[{\sc casa};][]{2007ASPC..376..127M} software package.  The
initial calibration steps applied the necessary baseline corrections
and removed any data affected by shadowing, radio frequency
interference (RFI) or instrumental problems.  We then used the primary
calibrator 3C\,286 to carry out bandpass and flux density calibration,
setting the flux scale according to the coefficients derived at the
EVLA by NRAO staff in 2010.  We used the secondary calibrator
J1707--1415 to derive amplitude and phase gains for \src.  After
applying the calibration, the data on \src\,were averaged in frequency
by a factor of 8, and then imaged using natural weighting.  The field
was not bright enough for self-calibration, so after a single round of
deconvolution, when detected, the source was fit in the image plane
with an elliptical Gaussian.  The fitting results are given in
Table~\ref{tab:evla}. We list the MJD times of these observations
  and the difference between the mid--times of the radio and X--ray
  exposures in Table~\ref{val-xrcorr}.

\begin{table*}
\caption{The values for the X--ray and radio flux for \src\, used in Figure~\ref{xrcorr}.}
\label{val-xrcorr}
\begin{center}
\begin{tabular}{cccccc}
\hline
Satellite & X--ray flux 1--10 keV & Radio flux  & Instrument & start MJD
X-ray obs & $|\Delta T_{X-radio}|^c$\\
    &  \flx     & mJy &  & days & days \\
\hline\hline
\swift & 2.4$\times 10^{-9}$ &  4.92$\pm$0.04 & WSRT  & 55465.7 & 0.14\\
\swift & 4.8$\times 10^{-9}$ &  $\approx$10 & EVN  &55469.7 & 0.42\\
\chan$^a$ & 1.73$\times 10^{-13}$ & 0.021$^b$ & EVLA & 55670.3$^a$ & 1.07$^a$\\
\chan & 8.3$\times 10^{-12}$ &  0.081$\pm$0.008 & EVLA & 55693.2 & 1.00\\
\swift & 1.4$\times 10^{-11}$ &  0.098$\pm$0.016 & EVLA  & 55721.2 & 6.01\\
\swift & 5.8$\times 10^{-12}$ &  0.102$\pm$0.012 & EVLA & 55741.0 & 0.12 \\
\chan & 4.5$\times 10^{-15}$ &  0.017$^b$ & EVLA  & 55788.8 & 0.57\\
\end{tabular}     
\end{center}
{\footnotesize $^a$ The average flux, combining the fluxes measured from 
Obs ID 12438 and 12439, is given and used in Figure~\ref{xrcorr}. The MJD 
is the average of the MJDs of Obs ID 12438 and 12439. The time
difference between the X--ray and radio observation is with respect to
this average.\\}
{\footnotesize $^b$ 3 $\sigma$ upper limit.\\}
{\footnotesize $^c$ Difference between the X-ray and radio
  mid-exposure times (except for footnote $a$ where the difference is
  with respect to the average of the start times of the two X--ray observations).\\}
\end{table*}

\begin{table*}
\begin{center}
{\caption{\label{tab:evla} EVLA observations of \src\, at 4.96
    GHz. The MJDs correspond to mid--exposure.}}
\begin{tabular}{cccccc}
\hline
Observation date & MJD & On source & Array & Bandwidth & Flux density \\
& (days; UTC) & exp.~time (min) &configuration &  (MHz) & (mJy\,beam$^{-1}$) \\
\hline\hline
2011 Apr.~19 & 55670.42$\pm0.04$ & 50 & B &  256 & $0.028^c$ \\
2011 Apr.~20$^a$ & 55671.37$\pm0.99$  & 2$\times$50 & B &  256 & $0.021^c$ \\
2011 Apr.~21 & 55672.33$\pm$0.04 & 50 & B &  256 & $0.029^c$ \\
2011 May 13  & 55694.31$\pm$0.08 & 117 & BnA &  256 &  $0.081\pm0.008$ \\
2011 Jun.~03 & 55715.21$\pm$0.04 & 50 & BnA &  256 &  $0.098\pm0.016$ \\
2011 Jun.~29 & 55741.14$\pm$0.04 & 50 & A &  256 &  $0.102\pm0.012$ \\
2011 Aug.~16& 55789.05$\pm$0.06 & 80 & A &  256 &  $0.024^c$ \\
2011 Aug.~16$^b$& 55789.58$\pm$0.59  & 2$\times$80 & A &  256 & $0.017^c$ \\
2011 Aug.~17& 55790.11$\pm$0.06  & 80 & A &  256 & $0.025^c$ \\
\hline
\end{tabular}
\end{center}
{\footnotesize $^a$ Average of the two 50 min.--long observations on April 19
  and 21. This value is used in Figure~\ref{xrcorr}.} \\
{\footnotesize $^b$ Average of the two 80 min.--long observations on August 16
  and 17. This value is used in Figure~\ref{xrcorr}.} \\
{\footnotesize $^c$ The 
$3\sigma$ upper limit to the source brightness is given.}\\
\end{table*}

\subsection{Radio -- X--ray correlation}
\label{correlation}
 
\begin{figure*} 
\includegraphics[angle=0,width=18cm,clip]{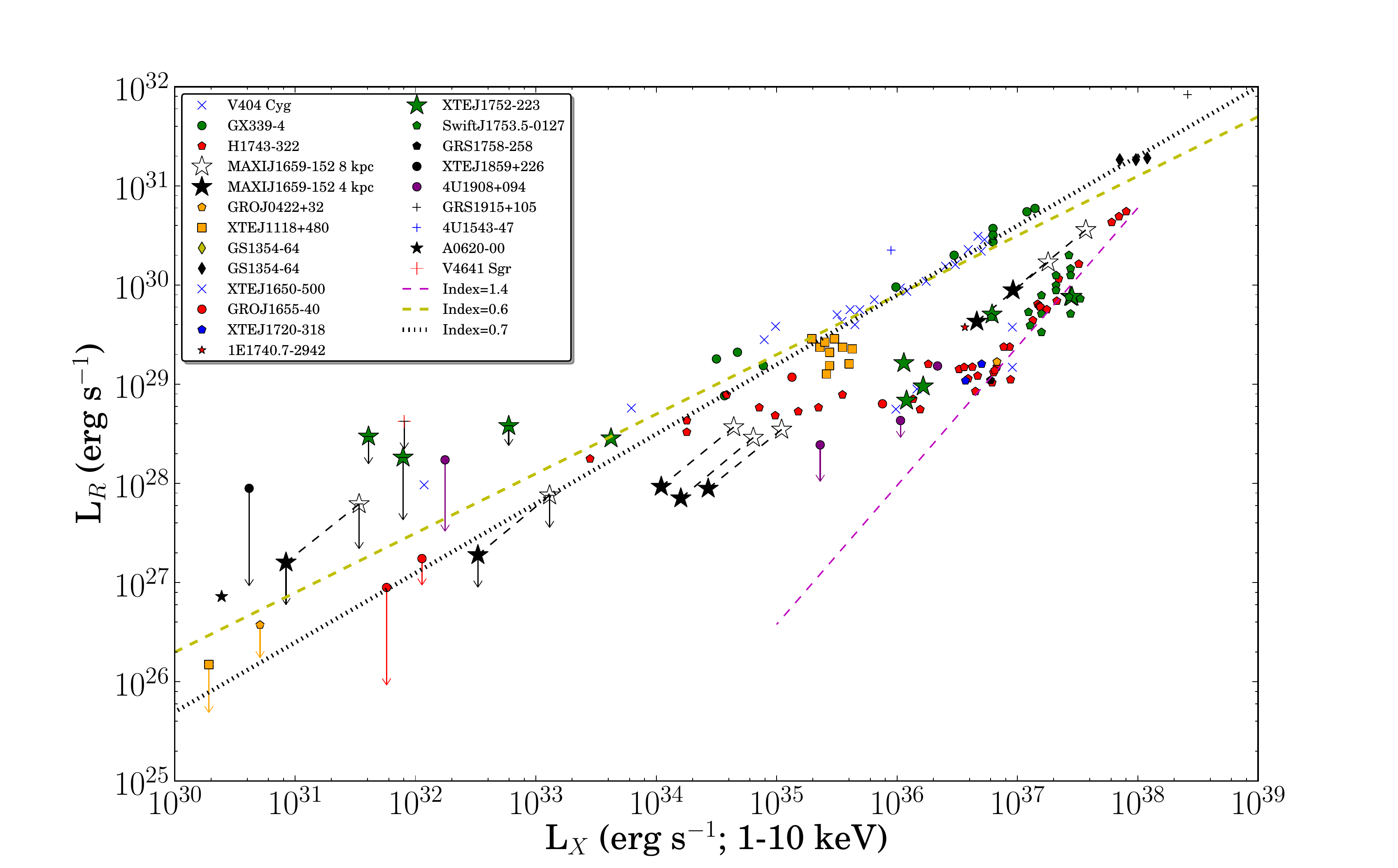}
\caption{The radio -- X--ray correlation including the points for
  \src\, using our quasi--simultaneous EVLA -- \chan/\swift\,
  observations. Furthermore, we used two quasi--simultaneous radio and
  X--ray observations from the low--hard state during the beginning of
  the outburst presented in the literature (see text for details). The
  unabsorbed 1--10 keV X--ray flux is converted to a luminosity using
  an estimated distance of 4 and 8 kpc for \src. The dashed green
  line traces the ${\rm L_R\propto L_x^{0.6}}$ correlation. The purple
  dashed line traces the ${\rm L_R\propto L_x^{1.4}}$ correlation and
  the dotted line traces a ${\rm L_R\propto L_x^{0.7}}$ correlation.
  The normalization for all these lines is arbitrary. Downward
  pointing arrows indicate upper limits to the radio luminosities. At
  X--ray luminosities above ${\rm L_X>1\times 10^{36}}$\lum, the
  points for \src\, fall below the correlation with power--law index
  0.6, indicating that \src\, follows the steeper correlation with
  power--law index of close to 1 (see \citealt{2012arXiv1203.4263G}). At
  intermediate luminosities (${\rm 1\times 10^{34}<L_X < 1\times
    10^{35}}$\lum) the points for \src\, make the transition in the
  direction of the 'standard' correlation with power--law index
  0.6. At the X--ray luminosity close to ${\rm L_X\approx 5\times
    10^{32}}$\lum\, the radio luminosity is too low for \src\, to be
  consistent with it being on the 'standard' correlation, unless the
  distance to the source is significantly more than 8 kpc, and/or, if
  the power--law index of the 'standard' correlation is steeper than
  0.6 for \src.}
\label{xrcorr} \end{figure*}

In Figure~\ref{xrcorr} we plot the observed correlation between the
X--ray and 4.96 GHz radio luminosities for \src\ using the X--ray
observations closest in time to the radio observations (converting the
fluxes in Table~\ref{val-xrcorr} to luminosities using distances of 4
and 8 kpc [see Discussion]). The quoted radio luminosities are
calculated starting from the measured flux density and assuming a flat
spectrum up to 5 GHz. Besides our quasi--simultaneous EVLA --
\chan/\swift\, observations we provide two data points using
quasi--simultaneous radio and \swift\, X--ray observations mentioned
in the literature. In particular, we used the radio detection obtained
with the Westerbork Synthesis Radio Telescope (WSRT) on Sept.~26, 2010
at 4.8 GHz reported in \citet{2010ATel.2874....1V} and the detection
using the European VLBI Network (EVN) at 4.9 GHz on Sept.~30, 2010
reported in \citet{2010ATel.2906....1P}. Both these observations were
early in the outburst of the source when it was possibly still
in the low--hard state (cf.~\citealt{2011ApJ...731L...2K}) 
  although the source spectrum showed continuous softening and \src\, may
  well have been in the hard--intermediate state at those times
  (\citealt{2011ApJ...731L...2K}; see Discussion). 

We derived the quasi--simultaneous 1-10 keV X--ray flux from the
measurements reported by \citet{2011ApJ...736...22K}. See
  Table~\ref{val-xrcorr} for the MJD times of these observations and
  the difference between the mid--times of the radio and X--ray
  exposures. We convert their 2--10 keV fluxes to 1--10 keV using the
webtool {\sc w3pimms} with the power--law index and interstellar
extinction, ${\rm N_H}$, provided in their paper as input for the {\sc
  w3pimms} tool. See Table~\ref{val-xrcorr} for all the flux values
that have been used to calculate the luminosities in
Figure~\ref{xrcorr}.

\section{Discussion}

We studied the source \src\, during the last phase of the 2010--2011
outburst using contemporaneous \chan/\swift\, and EVLA
observations. 

In order to optimally use the information gained from these
observations we first need a distance estimate for \src.
\citet{2011ApJ...736...22K} applied various scaling relations between
observables to obtain distance estimates, but they seem to settle on a
d$\approxgt 6.1$~kpc on the basis of the assumption that at the peak
of the outburst the source X--ray emission must be above
0.1$\times$\ledd.  \citet{2011ApJ...736...22K} mention that this
assumption contradicts the quiescent counterpart identified by
\citet{2010ATel.2976....1K}, as the putative quiescent optical
counterpart would be too bright given this distance and the known
orbital period. However, it only is too bright if there is negligible
light coming from the residual accretion disc in quiescence. In
several BHXRBs in quiescence, however, optical spectroscopy and
variability shows that the disc does contribute significantly to the
optical light on many occasions in quiescence
(cf.~\citealt{1994MNRAS.266..137M}, \citealt{2003ApJ...599.1254G};
\citealt{2010ApJ...710.1127C}). The star can of course be an
interloper star unrelated to \src: time--series optical spectroscopic
observations of the star while \src\, is in quiescence will be
necessary to decide on this issue.  \citet{2011arXiv1102.2102K} use
the relation between the outburst amplitude in the optical and the
orbital period between known sources and their (approximate) distances
to estimate a distance of 7$\pm$3 kpc for \src. In
\citet{2011ATel.3358....1M}, we noted that the reported relation
between the optical outburst amplitude and the orbital period
(\citealt{1998MNRAS.295L...1S}) does not take the source inclination
into account. A high inclination leaves a smaller projected surface
area of the accretion disc hence one would be inclined to put the
source artificially far away using this method.

Similarly, high inclination sources may be considerably fainter in
X--rays (cf.~\citealt{1988ApJ...324..995F};
\citealt{2005ApJ...623.1017N}) explaining the relatively low
luminosity at the peak of the outburst that one would derive for
distances lower than 6 kpc. Furthermore, the lack of eclipses in
sources with values $<$0.1 for the mass ratio, which we define here as
the mass of the secondary divided by the mass of the black hole
primary, can be accomodated for inclinations $<80^\circ$
(cf.~\citealt{1985MNRAS.213..129H}), this is significantly higher than
the canonical 60$^\circ$--75$^\circ$
(cf.~\citealt{2002apa..book.....F}) making the inclination dependent
effects on the X--ray luminosity larger.

Assuming that the accretion disc does not contribute significantly to
the optical light in quiescence and that the source found by
\citet{2010ATel.2976....1K} is related to \src\, we derived a distance
lower--limit between 1.6 and 4.2 kpc (\citealt{2011ATel.3358....1M}).
\citet{2012ApJ...746L..23K} obtained median-resolution optical
spectroscopy of \src\, during the outburst. Those authors use the
observed velocity shifts of the Na~I~D and Ca~II~H\&K interstellar
absorption lines with respect to that of the local standard of rest
and a kinematic model of Galactic rotation to derive a lower limit to
the source distance of 4$\pm 1$ kpc. These two lower limits together
seem to favour an M2~V companion star over an M5~V companion star as
the latter would lead to a lower limit of 1.6 kpc and could only be
consistent with a distance lower limit of 4 kpc if the accretion disc
is dominating the optical light. This leads us to adopt a lower limit
to the distance of \src\,of 4 kpc. Nevertheless, it has often been
found that the accretion disc contributes 10--45 per cent to the
optical $R$--band light in BHTs in quiescence (see references above
and e.g.~\citealt{2008MNRAS.387..788R} and references therein).
Conservatively assuming an accretion disc contribution of 50 per cent
to the $R$--band luminosity and a M2~V companion star we derive a
distance of 5.9 kpc, which we round to 6 kpc.  Taking uncertainties
such as that in the derived interstellar extinction, spectral type of
the companion star into account, and allowing for an even higher
accretion disc contribution, we subsequently assume that the distance
to \src\, is 6$\pm$2 kpc.

During the last two \chan\, observations the source flux is consistent
with being the same. We interpret this as evidence for the source
having reached quiescence, although a future \chan\, observation
will be necessary to test this better. For our estimated distance of 6$\pm$2 kpc
the source quiescent 0.5--10 keV luminosity is $1-4\times 10^{31}$\lum
for a 0.5--10 keV source flux of 5$\times 10^{-15}$\flx. The estimated
luminosity based on the orbital period -- X--ray luminosity
correlation would be $\approx 1-4\times 10^{30}$\lum, so significantly
lower than what we measured.

\begin{figure} 
\includegraphics[angle=270,width=8cm,clip]{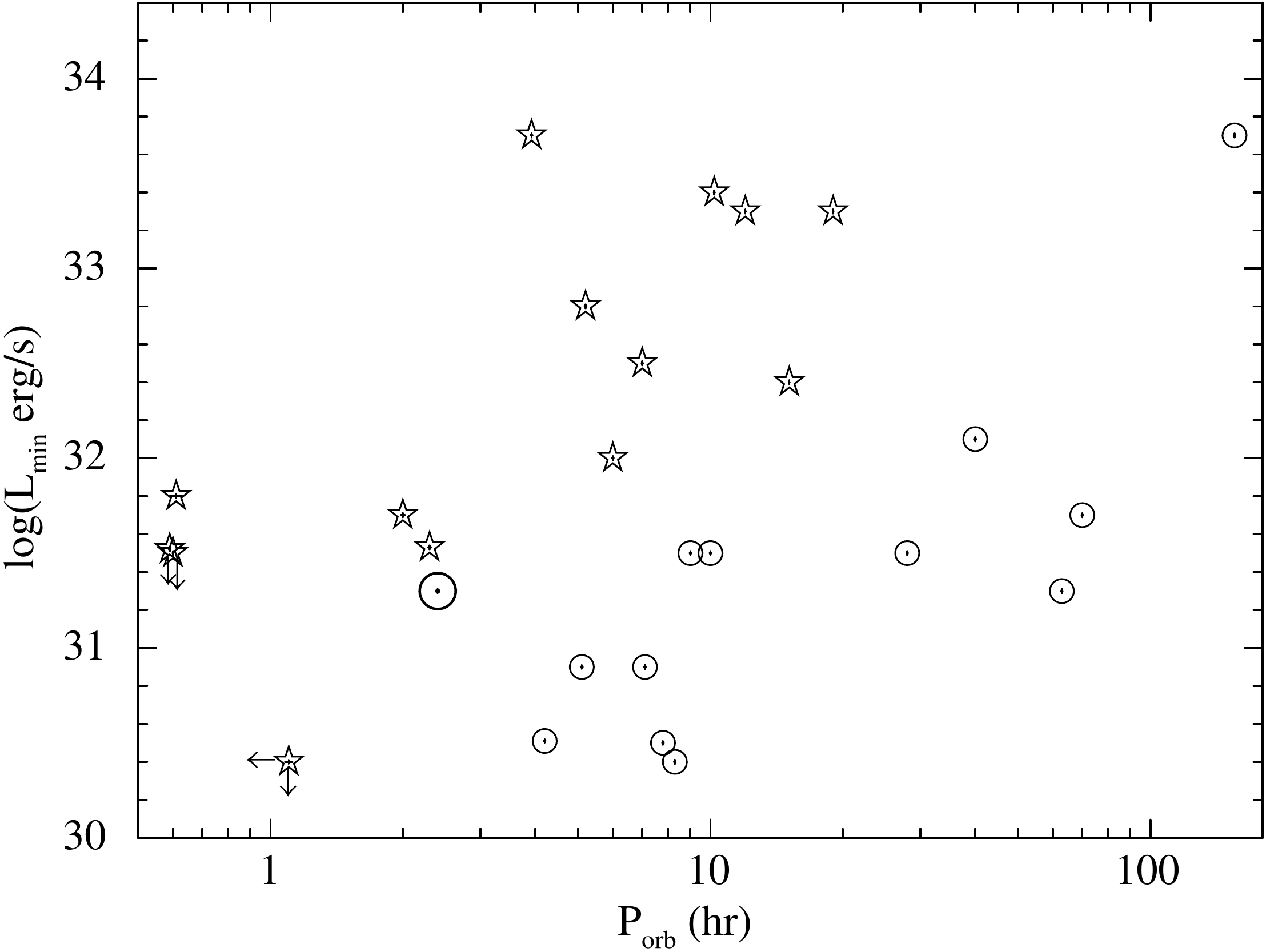}
\caption{The orbital period -- quiescent X--ray luminosity correlation
  including the point for \src\,(figure updated after
  \citealt{2011ApJ...729L..21R}).  The bullet with the largest circle
  around is \src\, assuming a distance of 6 kpc in calculating the
  X--ray luminosity. The star symbols are used for neutron stars
    and the encircled dots are used for black holes. Note that the
    uncertainties on the luminosities due to distance and flux
    measurement uncertainties (e.g.~\citealt{2004MNRAS.354..355J} for
    the distance uncertainties) are not included.}
\label{PorbLxcorr} \end{figure}

Hence, assuming the observed flux for the last two observations
  corresponds to the quiescent flux and taking the distance derived
  above, the source quiescent X--ray luminosity is on the high--end
for a black hole system given the short orbital period of 2.4~hr (see
Figure~\ref{PorbLxcorr}; cf.~\citealt{2001ApJ...553L..47G};
\citealt{2011ApJ...729L..21R}). The quiescent X--ray luminosity is
more in line with what would be expected for a neutron star at that
orbital period in quiescence. However, this would be at odds with the
observed spectral and timing properties as well as with the radio flux
during the outburst (\citealt{2011ApJ...736...22K};
\citealt{2011ApJ...731L...2K}; \citealt{2011MNRAS.415..292M}). If this
quiescent flux level is confirmed by future X--ray observations (an
additional deep \chan\, observation is planned for the Summer of 2012)
there is perhaps more scatter on the known correlation between
quiescent X--ray luminosity and orbital period, or the distance to the
source has been over--estimated. A distance of $\sim$2 kpc would be
required to bring the luminosity of \src\, to that estimated using the
orbital period -- X--ray luminosity relation. Possible effects of the
high inclination on the observed X--ray flux would make the intrinsic
luminosity higher still.

In order to study the relation between accretion and jet ejection in
\src\, we add points for \src\, to the known data on the radio --
X--ray correlation in BHXBs (Figure~\ref{xrcorr};
cf.~\citealt{2003A&A...400.1007C}; \citealt{2003MNRAS.344...60G};
\citealt{2004MNRAS.inpress}; \citealt{2010MNRAS.409..839C};
\citealt{2011MNRAS.414..677C}; \citealt{2011ApJ...739L..18M};
\citealt{2011MNRAS.413.2269S}).  From Figure~\ref{xrcorr} it seems
that \src\, follows the steeper correlation between the radio and
X--ray luminosity at high X--ray luminosities (${\rm L_X > 1\times
  10^{36}}$\lum) during the start of the outburst
(\citealt{2012arXiv1203.4263G}), although it is not certain that
  the two observations reflect the optically thick synchrotron
  emission typical for the low-hard state. For instance,
  \citet{2010ATel.2874....1V} report a linear polarization of 23 per
  cent for the WSRT observation on Sept.~26, 2010 which is
  significantly higher than the $\sim$10 per cent maximum for the
  linear polarisation for optically-thick synchrotron emission
  (\citealt{1994hea2.book.....L}). Furthermore, the source started a
  gradual transition to the soft state at the end of September 2010
  (\citealt{2011ApJ...731L...2K}; \citealt{2010ATel.2888....1K}),
  hence this possibly also affects the radio flux during the EVN
  observation on Sept.~30. In addition, the EVN observation may
  resolve out some of the emission. Finally, sub--millimeter
  observations at 23:30 UT on Sept.~25, 2010
  (\citealt{2010GCN..11304...1D}), just prior to the WSRT observation
  of Sept.~26, 2010 found the source at 12.6$\pm$2.4 mJy, a flux
  significantly higher than that at 4.8 GHz measured $\sim$15 hrs
  later with the WSRT. The high degree of linear polarisation together
  with the information on the sub--millimeter -- 4.8 GHz spectral
  energy distribution suggests that short--lived, transient, optically
  thin radio events contributed to the 4.8 GHz radio flux, similar to
  the GHz radio emission observed by \citet{2010MNRAS.401.1255J} in
  H~1743--322. 

During the decay of the outburst as traced by our EVLA and
\chan/\swift\, observations the source seems to make the transition
from the steep branch in the direction of the standard radio --
X--ray correlation with power--law index between 0.6 and 0.7, similar
to what has been observed by \citet{2011MNRAS.414..677C} for
H~1743--322 and by Ratti et al.~(2012; accepted) for XTE~J1752-223.
Although we only have information for two radio -- X--ray observations
when the source was brighter than $1\times 10^{36}$\lum\, those two
points have a lower X--ray luminosity than the bulk of sources in this
regime, or conversely, a higher radio luminosity, whereas the
correlation index seems to be close to the index of 1 of the sources in this branch
(\citealt{2012arXiv1203.4263G}).

At present there is no clear explanation for the steeper slope of
index $\sim 1$. Potentially, a value of $\sim 1$ for the power--law
index can be explained by optically thin radio emission and
radiatively inefficient emission of X--rays as both radio luminosity
and X--ray luminosity would scale as the mass accretion rate squared
in that case (cf.~\citealt{1995A&A...293..665F} and cf.~
\citealt{1995ApJ...452..710N}, respectively). However, if many
optically thin flares would be dominating the radio luminosity, one
would naively expect the sources to fall at higher luminosities than
those sources following the 'standard' correlation, the latter being a
minimum non--flaring track.

The fact that \src\, falls to the low X--ray luminosity side of the
main pack of sources following this steeper correlation can, however,
potentially be explained by the high inclination under which we view
\src. If the X--ray emission is formed close to the black hole and
accretion disc, its luminosity is reduced because of the high
inclination (\citealt{1988ApJ...324..995F} and
\citealt{2005ApJ...623.1017N}). The radio luminosity will not be
hindered by the (partial) self--obscuration and reduced aspect of the
source as the radio emission is thought to be formed further out than
the X--rays (here, we assume that the radio--jet axis is not nearly
parallel to the binary orbital plane). It is unclear if the bulk of
the X--ray emission is indeed formed close to the black hole, though.
This effect could also slightly alter, i.e.~lower, the index of the
radio -- X--ray correlation.  This would thus lead to
inclination--dependent scatter on the slope of the sources in the
radio -- X--ray correlation. Note that \citet{2011MNRAS.413.2269S}
investigated whether there is a relation between the correlation that
sources trace in the radio -- X--ray luminosity plane with inclination
but found none. This could mean that any effect of the inclination
would only become apparent in sources where the inclination angle is
very high. \src\, may be the only known example of such a very high
inclination so far.  Alternatively, the fact that \src\, has a higher
normalization than the main group of sources following the steeper
power--law index is $\sim$1 correlation could be due to a
higher--than--average radio luminosity potentially caused by the
presence of radio flares as we discussed above (see also
\citealt{2010MNRAS.401.1255J}).

Interestingly, at an X--ray luminosity of 3.3$\times
10^{32}$--1.3$\times 10^{33}$\lum\,\src\,was not detected at limits
well below the radio luminosity that would be expected extrapolating
the ${\rm L_R\propto L_X^{0.6}}$ correlation for GX~339--4 (see green
dashed line Figure~\ref{xrcorr}). Possibly the correlation index is
steeper for some sources (\citealt{2011ApJ...739L..18M} noted that the
best--fit to the data of A0620--00, V404~Cyg and GX~339--4 gives a
power--law index of 0.67).  Besides our measurements on \src\, other
sources may fall below the correlation with index 0.6 at low
luminosities as well (e.g.~the upper limit to the radio luminosity for
XTE~J1118+480; cf.~\citealt{2011ApJ...739L..18M}) although there is
uncertainty on the normalization due to distance uncertainties
(\citealt{2004MNRAS.354..355J}; \citealt{2004MNRAS.inpress}).
Nevertheless, our radio non--detection for an X--ray luminosity of
3.3$\times 10^{32}$\lum--1.3$\times 10^{33}$\lum, for a distance of 4
and 8 kpc, respectively is constraining. The discrepancy could be
alleviated if the source distance is more than 8 kpc, but this would
make the quiescent X--ray luminosity higher still. Only if our lowest
measured X--ray flux is not close to the true quiescent X--ray flux
and if the source is more than 10 kpc away is the source behaviour
consistent with that found for other BHTs before.

\section*{Acknowledgments} \noindent PGJ acknowledges support from a
VIDI grant from the Netherlands Organisation for Scientific
Research. Support for this work was provided by the National
Aeronautics and Space Administration through Chandra Award Number
GO0-11058A issued by the Chandra X-ray Observatory Center, which is
operated by the Smithsonian Astrophysical Observatory for and on
behalf of the National Aeronautics Space Administration under contract
NAS8-03060.  This research has made use of the SIMBAD database,
operated at CDS, Strasbourg, France, of NASA's Astrophysics Data
System Bibliographic Services, of SAOImage DS9, developed by
Smithsonian Astrophysical Observatory, of software provided by the
Chandra X-ray Center (CXC) in the application packages CIAO, and of
the XRT Data Analysis Software (XRTDAS) developed under the
responsibility of the ASI Science Data Center (ASDC), Italy.

\end{document}